\documentclass[12pt]{iopart}

\usepackage{float}
\usepackage{graphicx}

\begin{document}

\title[Role of viscoelasticity in the adhesion of mushroom-shaped pillars]{Role of viscoelasticity in the adhesion of mushroom-shaped pillars}

\author{Guido Violano, Savino Dibitonto and Luciano Afferrante}

\address{Department of Mechanics, Mathematics and Management, Polytechnic University of Bari, Via E. Orabona 4, Bari, 70125, Italy}
\ead{guido.violano@poliba.it}
\vspace{10pt}

\begin{abstract}
Mushroom-shaped pillars have been extensively studied for their superior adhesive properties, often drawing inspiration from natural attachment systems observed in insects. Typically, pillars are modeled with linear elastic materials in the literature; in reality, the soft materials used for their fabrication exhibit a rate-dependent constitutive behavior.

This study investigates the role of viscoelasticity in the adhesion between a mushroom-shaped pillar and a rigid flat countersurface. Interactions at the interface are assumed to be governed by van der Waals forces, and the material is modeled using a standard linear solid model. Normal push and release contact cycles are simulated at different approaching and retracting speeds.

Results reveal that, in the presence of an interfacial defect, a monotonically increasing trend in the pull-off force with pulling speed is observed, and the corresponding change in the contact pressure distribution suggests a transition from short-range to long-range adhesion. This phenomenon corroborates recent experimental and theoretical investigations. Moreover, the pull-off force remains invariant to the loading history, due to our assumption of a flat-flat contact interface. Conversely, in absence of defects, detachment occurs after reaching the theoretical contact strength, and the corresponding pull-off force is found to be rate independent.
\end{abstract}

\vspace{2pc}

\section{Introduction}

Numerous adhesive devices are designed to mimic the sticky performance of insect pads \cite{gorb2007}, whose textured surfaces significantly enhance adhesion \cite{lee2018,hensel2018}. Soft polymers with surfaces decorated by mushroom-shaped pillars are typically employed to manufacture bio-inspired adhesives \cite{gorb2007, lee2018, wang2014,zhang2015}. Experimental and theoretical studies \cite{gorb2007,bullock2011} have established that this geometry improves adhesive performance. Polymeric materials used in the manufacturing of adhesive devices exhibit viscoelastic properties, which are often overlooked in the modeling of the contact mechanics of adhesive pillars \cite{carbone2011}. Due to viscoelasticity, the contact problem becomes rate-dependent \cite{lorenz2013,violano2019B,violano2021rateA}, significantly complicating the modeling, even for the case of an isolated single pillar \cite{li2024}.

In recent years, there has been increasing interest in understanding adhesion between viscoelastic materials. For instance, Boundary Element Method (BEM) techniques have been developed to study 1D \cite{van2021, muser2022} and 2D contacts \cite{muller2023} between a viscoelastic substrate and a rigid indenter. However, in such investigations, the deformable viscoelastic solid is often simplified as a half-space, under the assumption that the contact area is relatively small compared to the bulk material volume. Finite Element Method (FEM), on the other hand, offers the advantage of modeling any geometry for the deformable solid \cite{li2024}, thereby avoiding the need for strong approximations inherent in half-space contacts \cite{violano2024}.

Although there have been numerous studies investigating the effect of pillars geometry on their adhesive behavior \cite{kim2020, carbone2013, aksak2014, micciche2014, paretkar2013}, research on the influence of viscoelasticity on the adhesive response of mushroom-shaped pillars remains sparse. Recently, Li et al. \cite{li2024} conducted experimental and FEM investigations into the pull-off dynamics of an individual mushroom-shaped pillar. Their findings revealed that, during the detachment process, there is a transition from an edge-crack mode to a center-crack mode with increasing retraction speed, particularly notable for mushrooms with an optimal cap thickness. It is known that the geometry of the mushroom cap significantly influences the behavior of interfacial cracks and the pull-off stress \cite{zhang2021,CarboneSMALL}. Furthermore, consistent with prior experimental studies on soft materials \cite{violano2021rateA, violano2021rateB}, the pull-off force was observed to increase with the retraction speed of the mushroom \cite{li2024}. This phenomenon is well-documented in the literature and attributed to viscoelasticity \cite{greenwood1981}. The material responds with a viscoelastic modulus that depends on the excitation frequency \cite{lorenz2013}. This leads to the emergence of an effective surface energy \cite{afferrante2022}, introduced to explain the rate-dependent adhesive behavior \cite{gent1972}.

Afferrante and Violano developed a finite element model to investigate the adhesion between a rigid Hertzian indenter and a viscoelastic substrate, incorporating dry adhesion with a finite-range potential \cite{afferrante2022}. This model has been recently expanded to examine the detachment modes of cylindrical and mushroom-shaped elastic pillars \cite{violMUSH}. Their research indicates that even in the absence of viscous dissipation, various detachment modes can occur during the debonding process due to the redistribution of interfacial pressures. In the present work, we extend our study by including viscous effects, investigating their impact on both the pull-off force and adhesive hysteresis. Moreover, our investigation encompasses the entire loading-unloading cycle and not only the debonding phase.

\section{Formulation of the problem}
\begin{figure}[H]
\centering\includegraphics[trim=5cm 3cm 6.8cm 3cm, clip, width=0.85\textwidth]{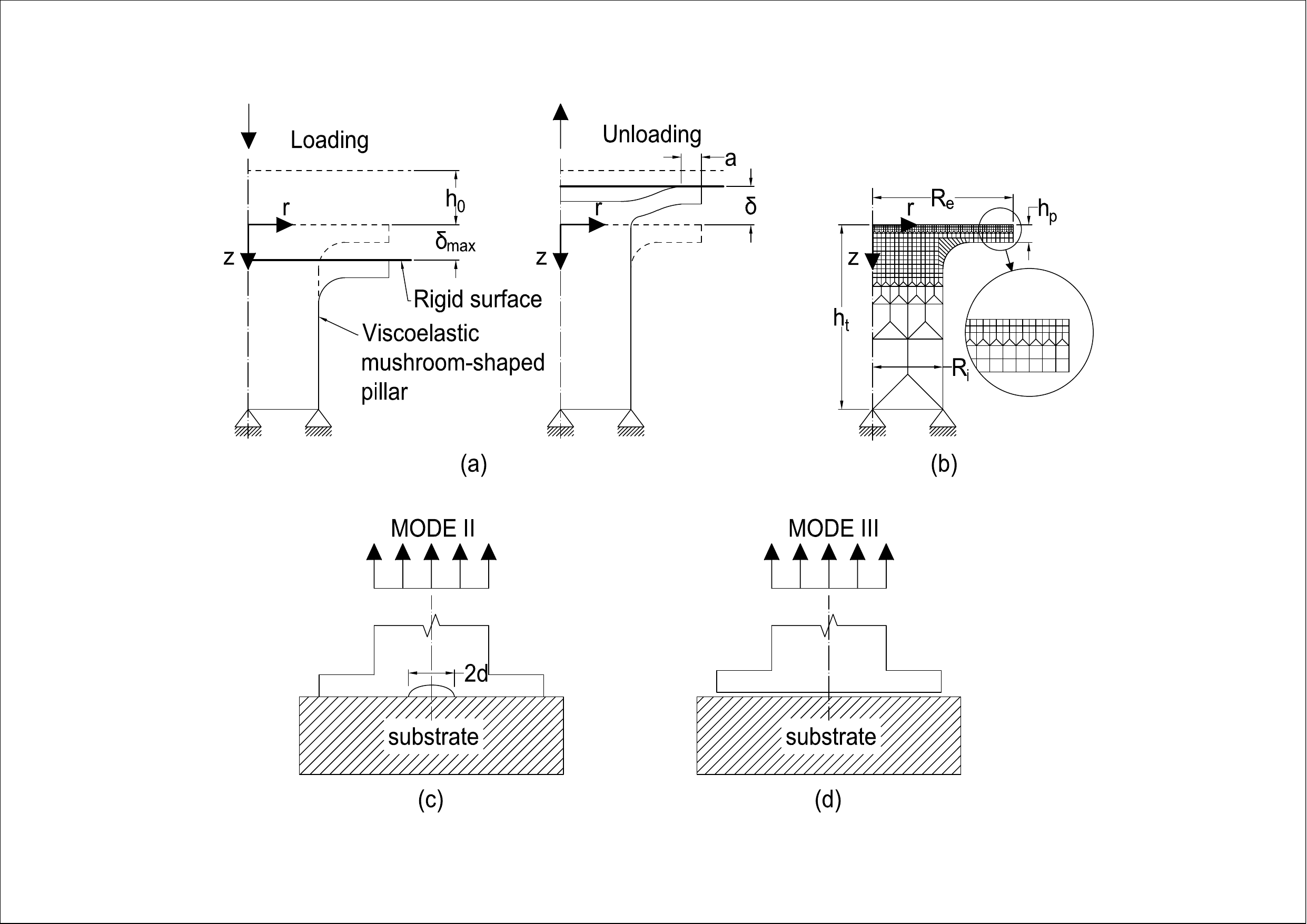}
    \caption{a) The problem under investigation: a rigid flat surface is pressed against a mushroom-shaped pillar at a constant speed $V$ and then pulled apart from it. b) A schematic representation of the FE model used for modeling the axisymmetric pillar. c) Mode II detachment in presence of an internal defect at the interface. d) Mode III detachment in absence of interfacial defects.}
    \label{PROBLEM}
\end{figure}

Figure \ref{PROBLEM}a shows the problem under investigation: a rigid flat surface is pressed into contact against a mushroom-shaped pillar and then pulled apart from it. The methodology used for this model follows the approach described in Refs. \cite{afferrante2022, violano2022A, afferrante2023B}, where further details are provided.

The pillar is modeled using linear axisymmetric elements and is constrained at its base (figure \ref{PROBLEM}b). The rigid flat surface is controlled via a single master node to which the displacement is applied. The viscoelastic modulus $E(t)$ of the substrate is given by the linear standard model in its Maxwell representation

\begin{equation}
E(t)=E_0 + (E_{\infty} - E_0) \exp (-t/\tau)
\end{equation}%
being $E_0$ and $E_{\infty}$ the relaxed and instantaneous values of the viscoelastic modulus, respectively, and $\tau$ is the relaxation time. 

To replicate the adhesive interactions between the pillar and the countersurface, we build nonlinear elements positioned at the contact interface. The constitutive law of these elements is described through the traction-displacement relationship derived from the Lennard-Jones (LJ) potential law

\begin{equation}
\sigma_{\mathrm{LJ}} \left( r\right) =\frac{8\Delta \gamma }{3\epsilon }\left[ \left( 
\frac{\epsilon }{g\left( r\right) }\right) ^{3}-\left( \frac{\epsilon 
}{g\left( r\right) }\right) ^{9}\right]
\label{LJ}
\end{equation}%
where $\Delta \gamma $ is the 'quasi-static' surface energy of adhesion, $g\left( r\right) $ is the interfacial gap and $\epsilon$ is the range of action of the adhesive forces.

Carbone et al. \cite{carbone2011} investigated the three potential detachment modes for cylindrical and mushroom-shaped pillars, identifying three detachment modes. The first, Mode I, is characterized by crack propagating from the contact edge; the second, Mode II, occurs when propagation originates from an interfacial defect; the last, Mode III, is the detachment for decohesion of the interface. It occurs at the theoretical contact strength, that, for the interaction law (\ref{LJ}), is $16 \Delta\gamma/(9 \sqrt{3} \epsilon)$.

The presence of a plate at the top of a cylindrical pillar eliminates the stress singularity at the outer edge, thereby preventing the onset of Mode I detachment \cite{carbone2011}. Consequently, the only possible detachment mechanisms for a mushroom-shaped pillar are Mode II (Fig. \ref{PROBLEM}c) and Mode III (Fig. \ref{PROBLEM}d). 
To simulate the presence of a defect of length $2d$ at the interface, we remove adhesive elements from the center of the pillar for an equivalent length. This scenario can realistically occur in pillars that exhibit radially varying mechanical and adhesive properties \cite{kossa2023}, or for deposition of dust particles at the interface \cite{carbone2011}.

\section{Results}
The scope of our study is to investigate the impact of rate-dependent effects stemming from the intrinsic viscoelasticity of the pillar material. For this reason, approach-retraction contact cycles at different driving speeds are conducted. The results are presented in terms of dimensionless quantities: load $\hat{F} = F/(E_{0}^{*}R_{\mathrm{i}}^{2})$, contact radius $\hat{a} = a/R_{\mathrm{i}}$, penetration $\hat{\delta} = \delta/\epsilon$, contact pressure $\hat{\sigma} = \sigma \times (9 \sqrt{3}\epsilon)/(16\Delta\gamma)$, driving speed $\hat{V}=V \tau/\epsilon$, and defect size $\hat{d} = d/R_{\mathrm{i}}$.

The quantity $E_{0}^{*} = E_{0}/(1-\nu^2)$ is the plain strain elastic modulus of the pillar, with $\nu = 0.49$ being the Poisson ratio. Plots are obtained for $\Delta\gamma/(E_{0}^{*} \epsilon) = 0.02$, total height of the pillar $\hat{h}_\mathrm{t} = h_\mathrm{t}/R_{\mathrm{i}} = 2.38$, and $R_{\mathrm{i}} = 0.5 \, \mathrm{\mu m}$. The external dimensionless radius and plate thickness are assumed to be $\hat{R}_{\mathrm{e}} = 2$ and $\hat{h}_\mathrm{p} = 0.22$, respectively, while the fillet radius between pillar and plate is approximately $\hat{\rho} = r/R_{\mathrm{i}} = 0.22$. These values are consistent with those of real textured adhesive surfaces \cite{aksak2011RATIO, delcampo2007ALTEZZA}. 

All results are given for a viscoelastic materials with a Maxwell constitutive law with $E_{\infty}/E_{0} = 10$ and $\tau = 10^{-4}$ s.

\subsection{Pillar with an interfacial defect}
The presence of an internal defect triggers Mode II detachment \cite{carbone2011,violMUSH}, with the crack propagating from the center towards the outer edge of the contact interface.
Calculations are performed for a defect size $2\hat{d}=1$.

\subsubsection{Unloading from the relaxed state}
In the first set of simulations, the rigid indenter is brought into contact with the mushroom-shaped pillar at a very low driving speed $\hat{V} = 10^{-5}$, ensuring that the viscoelastic material response remains within its rubbery region, thereby preventing viscous dissipation. Consequently, the retraction phase always begins from a fully relaxed state of the material.

\begin{figure}[H]
    \centering\includegraphics[scale=0.65]{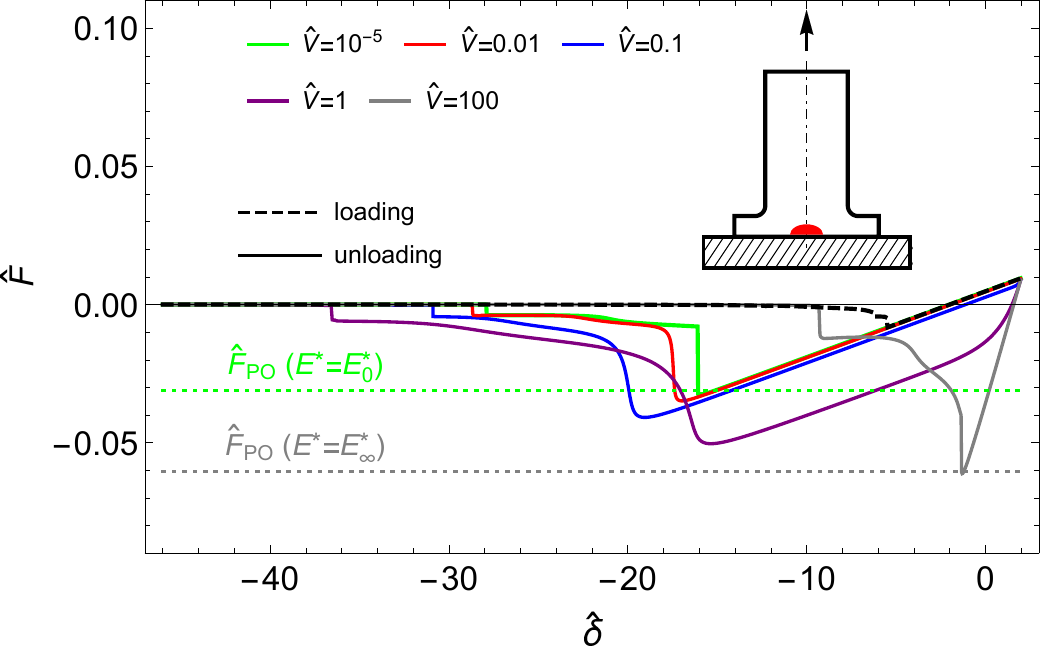}
    \caption{Dimensionless force $\hat{F}$ as a function of the dimensionless penetration $\hat{\delta}$ during the approach (black dashed line) and retraction (colored solid lines) phases. Retraction starts from the relaxed state of the viscoelastic material. The results are obtained for a pillar with an interfacial defect of size $2\hat{d}=1$.}
    \label{F2}
\end{figure}

Figure \ref{F2} shows the dimensionless force $\hat{F}$ as a function of the dimensionless penetration $\hat{\delta}$ during both the approach (black dashed line) and retraction (colored solid lines) phases. Unloading curves are obtained at different driving speeds $\hat{V}$. Contact hysteresis, given by the area enclosed in a contact cycle, occurs even when viscous dissipation is negligible (i.e., at 'small' or 'high' driving speeds). In these cases, hysteresis is strictly related to adhesive instabilities occurring at jump-in and jump-out of contact \cite{violano2021HYST, wang2021}. At intermediate speeds, i.e., when the material response is within its transition region, hysteresis results from the combination of adhesion instabilities and viscous effects \cite{muller2023,violano2021rateB}. In the specific case under investigation, detachment is expected to occur in mode II \cite{violMUSH} with crack moving from the inner to the outer edge of the contact interface. According to theoretical predictions \cite{carbone2013}, detachment is triggered when the average interfacial stress reaches $\sigma_{\mathrm{II}} = \sqrt{\pi E^{*} \Delta \gamma /(2d)}$. The green and gray dotted lines show the theoretical detachment force $F_{\mathrm{PO}}$, with $E^{*} = E_{0}^{*}$ at low retraction speeds and $E^{*} = E_{\infty}^{*}$ at high retraction speeds. However, agreement with the numerical pull-off force is obtained when the theoretical detachment force is calculated as $\sigma_{\mathrm{II}} \pi a_{\mathrm{PO}}^{2}$, where $\pi a_{\mathrm{PO}}^{2} = \pi (R_{\mathrm{i}}^{2} - d_{\mathrm{PO}}^{2})$ is the contact area at the moment of pull-off. A non-negligible difference occurs when, in the calculation of $F_{\mathrm{PO}}$, we adopt the initial contact area $\pi a^{2} = \pi (R_{\mathrm{i}}^{2} - d^{2})$ as suggested by the theoretical approach \cite{carbone2013}. This finding indicates that accurately estimating the detachment force requires accounting for the entire unloading process, as also suggested in Ref. \cite{violMUSH}. Theoretical predictions alone \cite{carbone2013} can only indicate the potential detachment mode.

\begin{figure}[H]
    \centering\includegraphics[scale=0.65]{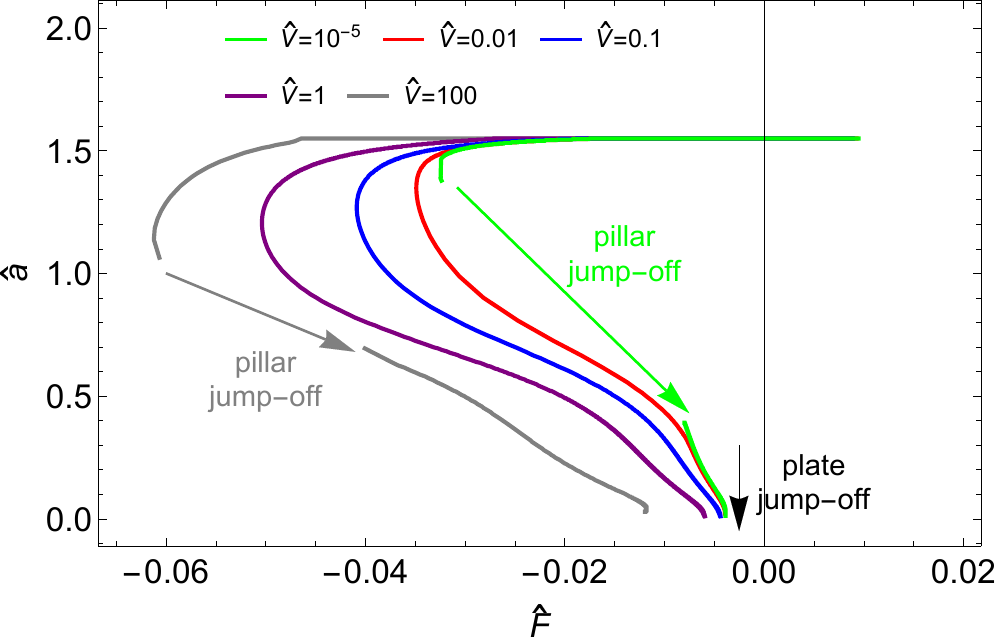}
    \caption{Dimensionless contact radius $\hat{a}$ as a function of the dimensionless force $\hat{F}$. The results are obtained during the retraction phase, which always starts from the relaxed state of the viscoelastic material, and for a pillar with an interfacial defect of size $2\hat{d}=1$.}
    \label{aF2}
\end{figure}

Figure \ref{aF2} shows the contact radius $\hat{a}$ as a function of the applied load during the retraction phase. At both 'small' and 'high' speeds, the detachment process is characterized by two distinct jumps, corresponding to the pillar jump-off and plate jump-off, respectively \cite{violMUSH}. These unstable jumps are highlighted in the plot with arrows. Conversely, at intermediate speeds, the detachment process is always continuous, demonstrating that viscoelasticity dampens contact instabilities. This behavior is similar to what is observed in viscoelastic rough contacts \cite{afferrante2023B}.

\begin{figure}[H]
    \centering\includegraphics[scale=0.65]{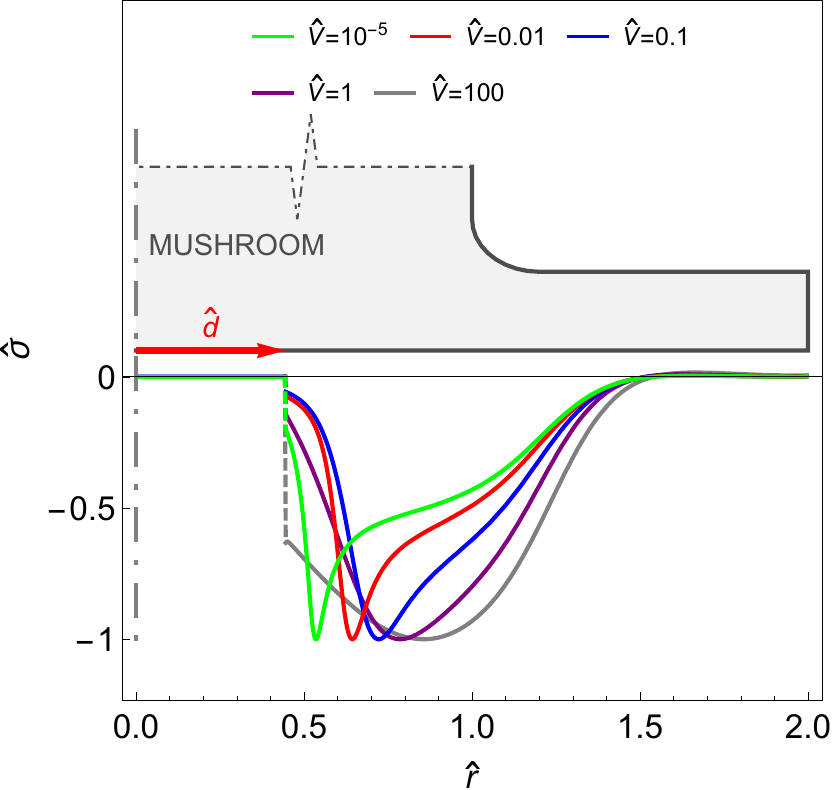}
    \caption{Interfacial pressure distribution $\hat{\sigma}$ at the pull-off point for different unloading speeds. The results are obtained for a pillar with an interfacial defect of size $2\hat{d}=1$.}
    \label{pr2}
\end{figure}

Figure \ref{pr2} shows the interfacial pressure distribution at the pull-off point, where the maximum tensile (negative) load is reached. As the speed increases from small to high values, the smoothing of the pressure peak indicates a transition from short-range to long-range adhesion \cite{li2024,violanoRANGE}. This transition occurs because, at high speeds, the material behaves more like a stiffer elastic medium, leading to a shift from JKR-like to DMT-like adhesion \cite{maugis1992}.

\subsubsection{Unloading from an unrelaxed state}
In many real applications, unloading does not start from a fully relaxed state of the pillar's bulk material. For example, in pick-and-place tools \cite{meitl2006}, the operating speeds during loading-unloading cycles do not allow the viscoelastic material to relax completely. In such cases, the material response is expected to be different and, therefore, warrants further investigation.

In the following set of simulations, the rigid flat indenter is approached and then retracted from the pillar at a constant driving speed $\hat{V}$, without any dwell time between the two phases. In this scenario, the material cannot relax before unloading.

\begin{figure}[H]
    \centering\includegraphics[scale=0.65]{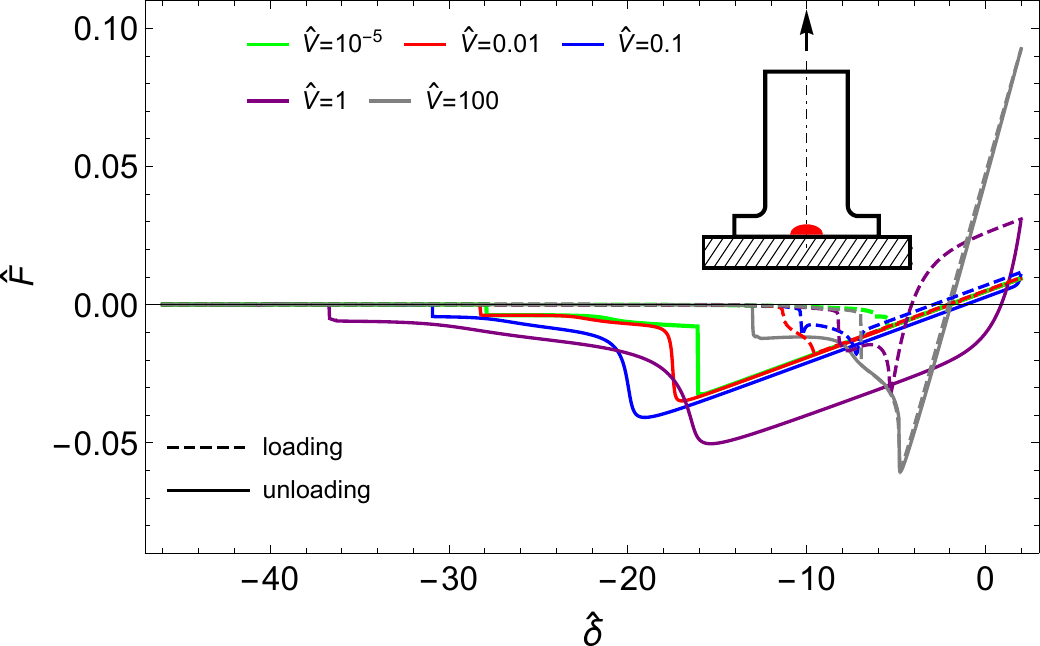}
    \caption{Dimensionless force $\hat{F}$ plotted against the dimensionless penetration $\hat{\delta}$ during the approach (dashed lines) and retraction (solid lines) phases. Retraction begins immediately after the approach. The results are obtained for a pillar with an interfacial defect of size $2\hat{d}=1$.}
    \label{F1}
\end{figure}

Figure \ref{F1} illustrates the dimensionless force $\hat{F}$ as a function of the dimensionless penetration $\hat{\delta}$. Loading and unloading data are represented by dashed and solid lines, respectively. While the approach and retracting curves are significantly influenced by the value of $\hat{V}$, it is noteworthy that the pull-off force is only marginally affected by the loading history, consistent with the findings of Ref. \cite{papangelo2023}. Conversely, for non-conformal contacts, such as the Hertzian one, we know the pull-off force is strongly influenced by the loading history \cite{violano2022A}.

\begin{figure}[H]
    \centering\includegraphics[scale=0.65]{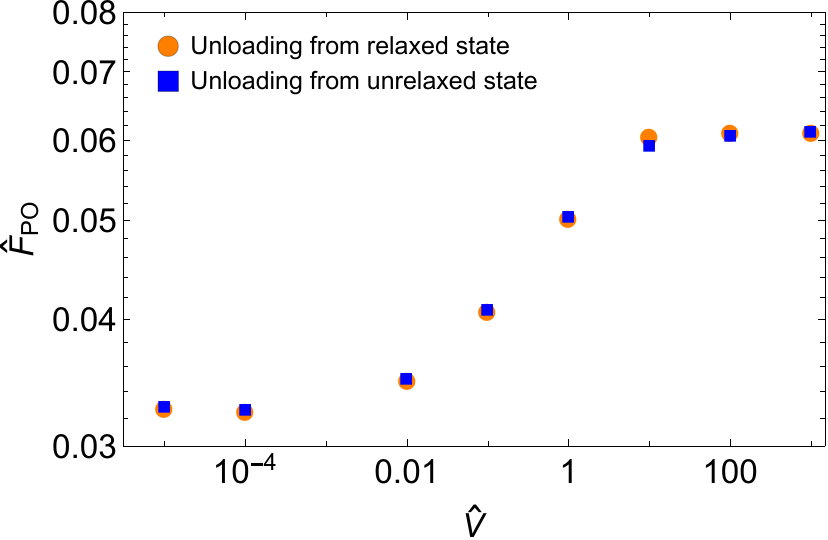}
    \caption{Dimensionless pull-off force $\hat{F}_{\textrm{PO}}$ as a function of the dimensionless pulling speed $\hat{V}$. The results are obtained for a pillar with an interfacial defect of size $2\hat{d}=1$.}
    \label{FPO2}
\end{figure}

Figure \ref{FPO2} shows the variation of the pull-off force with the retraction speed, when unloading is performed both from a relaxed and an unrelaxed state of the material. As anticipated, no differences are discernible. Moreover, the pull-off force increases with the pulling speed, reaching a plateau at higher speeds, in line with findings from previous studies on viscoelastic adhesive contacts \cite{muser2022,afferrante2022,PB2005,Greenwood2004}.

\subsection{Pillar without defect at the interface}
In the absence of an interfacial defect, the pillar is expected detaching consistently via Mode III, involving decohesion after reaching the theoretical contact strength \cite{violMUSH}.

\begin{figure}[H]
    \centering\includegraphics[scale=0.65]{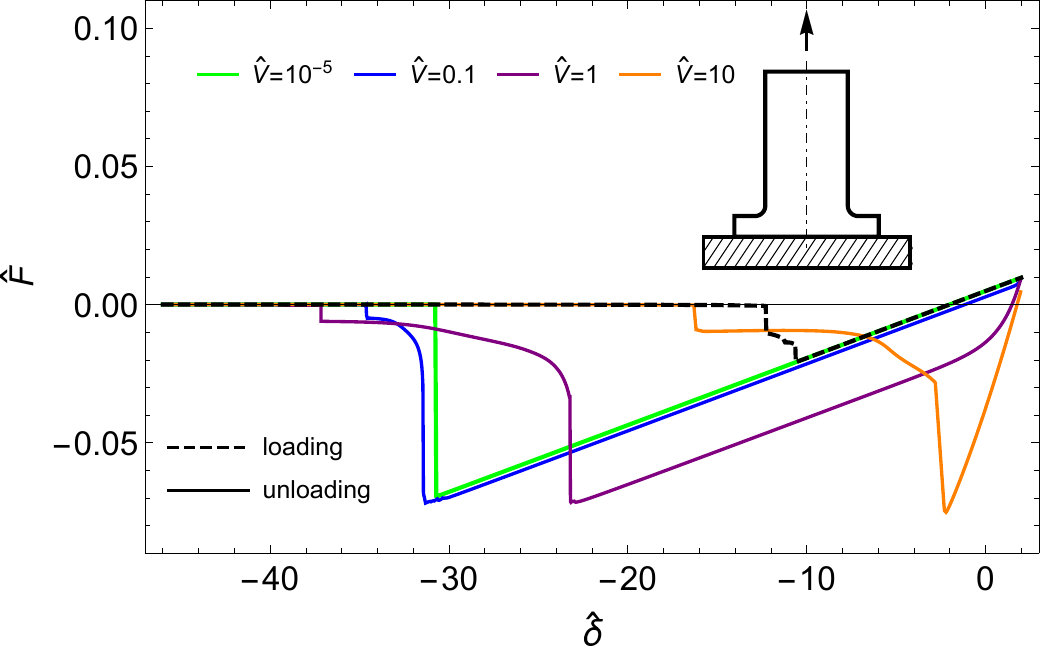}
    \caption{Normalized force $\hat{F}$ plotted against the dimensionless penetration $\hat{\delta}$ during the approach (dashed lines) and retraction (solid lines) of the flat indenter. Retraction starts from the relaxed state of the viscoelastic material and in absence of an interfacial defect.}
    \label{F3}
\end{figure}

As observed in the previous section, the retraction phase shows minimal sensitivity to the loading history. Therefore, in the subsequent analysis, we will focus on unloading from the relaxed state of the viscoelastic material. For this purpose, the approach phase is conducted at a 'low' speed. However, retraction is carried out at various pulling speeds.

\begin{figure}[H]
    \centering\includegraphics[scale=0.65]{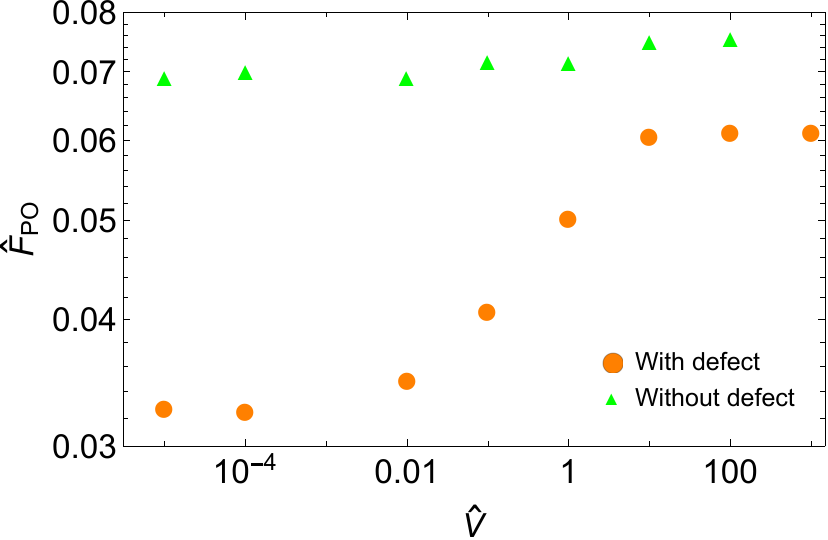}
    \caption{Adimensional pull-off force $\hat{F}_{\textrm{PO}}$ as a function of the pulling speed $\hat{V}$ for the same pillar with and without interfacial defect.}
    \label{FPO3}
\end{figure}

\begin{figure}[H]
    \centering\includegraphics[scale=0.65]{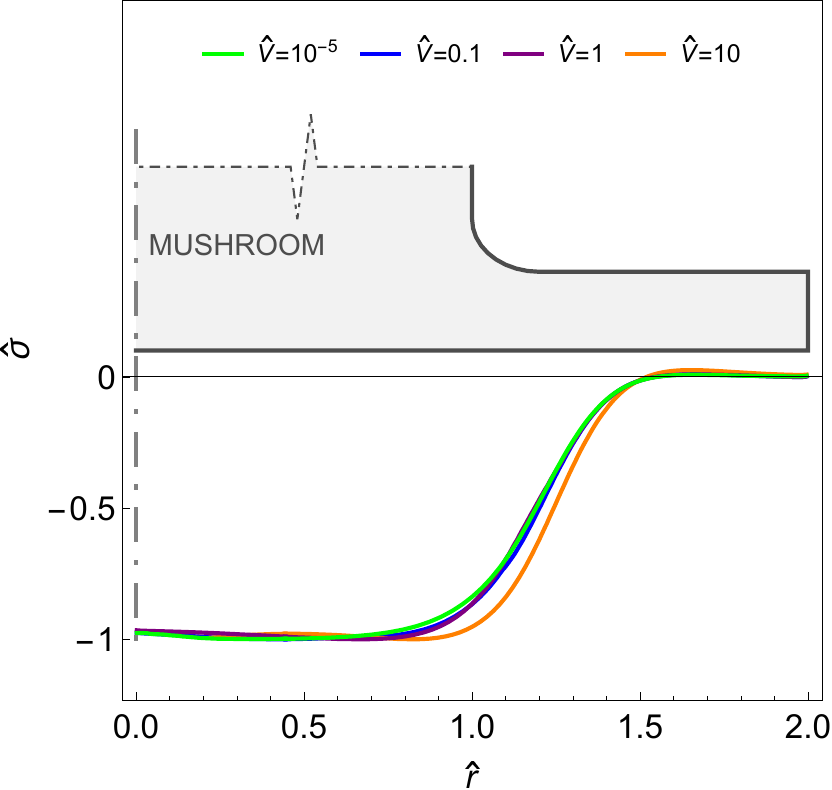}
    \caption{Interfacial pressure distribution $\hat{\sigma}$ at the pull-off point for a pillar without interfacial defect.}
    \label{pr3}
\end{figure}

Figure \ref{F3} illustrates the force-penetration curves obtained for the same approach speed (black dashed line) and different pulling speeds (solid colored lines). Despite the qualitative differences among the unloading curves, the pull-off force remains almost constant, regardless of the value of $\hat{V}$. This is further confirmed in Figure \ref{FPO3}, which plots the dependence of the pull-off force on the speed for the two scenarios considered in our study: the presence or absence of an interfacial defect. In the latter case (green triangles), the debonding force is clearly unaffected by the unloading speed, indicating that rate-dependent viscous effects do not alter the effective surface energy. 

In this framework, with detachment occurring via Mode III and independently of the deformation rate, the contact pressure at the pull-off point is uniformly distributed beneath the pillar. It reaches its maximum value, which corresponds to the theoretical strength $16 \Delta\gamma/(9 \sqrt{3} \epsilon)$ (see Figure \ref{pr3}), and exhibits negligible sensitivity to the pulling speed.

\section{Conclusions}
The modeling of bio-inspired mushroom-shaped pillars has traditionally overlooked the viscoelastic properties of the materials employed in these adhesive devices \cite{carbone2011, CarboneSMALL}. However, our research highlights the crucial role of viscoelasticity in shaping the adhesive behavior of mushroom-shaped pillars. Specifically, the presence of a defect at the interface makes the pull-off force strongly dependent on the retraction speed, although it reaches a maximum plateau at high pulling speeds.

Moreover, results indicate that the pull-off force is independent of loading history when the contact interface is assumed to be flat \cite{papangelo2023}. Nonetheless, we anticipate deviations from this observation upon the introduction of surface roughness at the interface \cite{afferrante2023B}, or in scenarios where one or both contacting surfaces exhibit macroscopic curvature, such as Hertzian contacts \cite{violano2022A}. Additionally, our results suggest that the pull-off force remains unaffected by the pulling rate during decohesion-mediated detachment.\\

\bibliographystyle{elsarticle-num}

\end{document}